\date{}
\documentclass[aps,pra,showpacs,twocolumn]{revtex4-1}
\usepackage{graphicx,psfrag,amsmath,amssymb,amsfonts,latexsym,color,dcolumn}
\begin{document}
\title{Fourth order perturbative expansion for the Casimir energy with a
slightly deformed plate}
\author{C\'esar D. Fosco$^{1,2}$}
\author{Fernando C. Lombardo$^3$}
\author{Francisco D. Mazzitelli$^{1}$}

\affiliation{$^1$ Centro At\'omico Bariloche,
Comisi\'on Nacional de Energ\'\i a At\'omica,
R8402AGP Bariloche, Argentina}
\affiliation{$^2$ Instituto Balseiro,
Universidad Nacional de Cuyo,
R8402AGP Bariloche, Argentina}
\affiliation{$^3$ Departamento de F\'\i sica {\it Juan Jos\'e
 Giambiagi}, FCEyN UBA, Facultad de Ciencias Exactas y Naturales,
 Ciudad Universitaria, Pabell\' on I, 1428 Buenos Aires, Argentina - IFIBA}
\date{today}
\begin{abstract} 
We apply a perturbative approach to evaluate the Casimir energy for a
massless real scalar field in $3+1$ dimensions, subject to Dirichlet
boundary conditions on two surfaces. One of the surfaces is assumed to be
flat, while the other corresponds to a small deformation, described by a
single function $\eta$, of a flat mirror.  The perturbative expansion is
carried out up to the fourth order in the deformation $\eta$, and the
results are applied to the calculation of the Casimir energy for corrugated
mirrors in front of a plane. We also reconsider the proximity force
approximation within the context of this expansion. 
\end{abstract}
\maketitle

\section{Introduction}\label{sec:intro}
Among the different approximate analytical approaches to the calculation of
the Casimir energy and related objects~\cite{libros}, a particularly
interesting one has been developed some time ago \cite{perturb}, for cases
where the mirrors' configurations can be described as a small geometrical
deformation of the simplest possible case: two infinite, flat, parallel
mirrors.  We shall focus here, in particular, on cases where just one of
the surfaces defined by the (zero width) mirrors is deformed.  Since one of
the surfaces is, indeed, flat, it is always possible to adopt Cartesian
coordinates $x_1,x_2,x_3$ in such a way that surface coincides with
the $x_3=0$ plane. The other mirror's surface will, in turn, be assumed to
be a small deformation of the $x_3=a$ ($a >0$) plane, namely: \mbox{$x_3
\,=\, a +\eta({\mathbf x_\parallel})$}.  Here, $\eta$ is a scalar function
of \mbox{${\mathbf x_\parallel}\equiv (x_1,x_2)$}, coordinates which are
then parallel to the flat mirror surface.  Hence, $\eta$ represents the
(assumed) small departure from a configuration corresponding to two flat
parallel mirrors (i.e., $x_3=0,a$).  

As already mentioned, for this, and similar situations, a perturbative
expansion in powers of $\eta({\mathbf x_\parallel})$ has been developed, in
order to compute the Casimir energy, under the assumption that $|\eta
({\mathbf x_\parallel})|< < a$.  This program has been carried out up to
the second order in $\eta$ \cite{perturb}, and applied, for example, to the
evaluation of the effect of corrugations on the Casimir
energy~\cite{perturb, perturb2,schaden}.  The perturbative 
calculations of Refs. 
\cite{perturb,schaden} are based on a functional approach, that we will also follow here.
On the other hand, in Ref. \cite{perturb2} a scattering approach is used,
and a general formula for the Casimir energy is given in terms of
the reflection coefficients associated to the mirrors. In this approach, one can incorporate 
finite temperature, finite conductivity and roughness corrections. However, as far as we 
know, all previous calculations for the normal force \cite{perturb, perturb2,schaden} have been performed up to 
second order in $\eta$.
  
In recent years, accurate numerical evaluations of the Casimir energy for
several geometries, including corrugated surfaces of arbitrary amplitude,
have become available. In spite of that, approximate analytical results are
always welcome, for a variety of reasons. Among them, to improve the
physical understanding of the dependence of the Casimir force with the
geometry, as well as to use them as benchmarks of complex numerical
calculations. 

It is the aim of this paper to present new analytical results for the
perturbative expansion, which had been implemented in previous works up to the second order
in $\eta$, by calculating the two subsequent corrections, of order three  and
fourth in the same expansion (note that a  similar program has been carried out 
in Ref. \cite{milton},  to compute the {\it lateral} force between corrugated surfaces
up to the fourth order in the amplitude).
  
The structure of this paper is as follows: in Sec.~\ref{sec:thesys}, we
introduce the system considered and derive the (formal) expression for the
Casimir energy, in terms of an ``effective action'', using a functional approach. 
In Sec.~\ref{sec:expansion}, we introduce the perturbative expansion for
the effective action, focusing on the expressions corresponding to terms of up to the
fourth order.  In Sec.~\ref{sec:results}, we present the explicit results up to 
the fourth order term. In Sec.~\ref{sec:other} we analyze further
approximations and resummations of the perturbative expansion. In
particular, we will see that the proximity force approximation (PFA) can be
understood as a resummation  of the perturbative expansion when the form
factors involved in the expansion are replaced by their zero-momentum
values. In Sec.~\ref{sec:examples} we apply the results to  compute the
correction to the Casimir energy up to the fourth order for a sinusoidally
corrugated surface,  and up to the third order for a corrugation described
by a combination of two sinusoidal functions of different wavelengths.
 We summarize our findings in Sec.~\ref{sec:conc}.
\section{The system}\label{sec:thesys}
We adopt Euclidean conventions, whereby spacetime coordinates are denoted
by $x^\mu \equiv x_\mu$ \mbox{($\mu=0,1,2,3$)}, $x_0$ being the imaginary
time and $x_i$, \mbox{($i=1,2,3$)} the spatial coordinates.  We shall also
use the notation \mbox{$x_\parallel \equiv (x_0,x_1,x_2)$} \mbox{$\equiv
(x_0,{\mathbf x_\parallel})$}, which shall be used as spacetime coordinates
on each mirror.

As mentioned above, the scalar field $\varphi$ satisfies Dirichlet
boundary conditions on two surfaces: $L$, which is the $x_3=0$ plane, while
the other, $R$, can be represented by a single function of 
${\mathbf x_\parallel}$:
\begin{eqnarray}\label{eq:deflr}
L) \;\; x_3 &=& 0 \nonumber\\
	R) \;\; x_3 &=& a + \eta({\mathbf x_\parallel}) \;.
\end{eqnarray}

Following the functional approach to the Casimir effect, we introduce
${\mathcal Z}$, the zero temperature limit of the partition function for
$\varphi$ in the presence of the two mirrors:
\begin{equation}
	{\mathcal Z}\;\equiv\; \int {\mathcal D}\varphi \;\delta_L(\varphi) \,
	\delta_R(\varphi) \; e^{- S_0(\varphi) } \;,
\end{equation}
where $S_0$ is the massless real scalar field Euclidean action
\begin{equation}
S_0(\varphi) \;=\; \frac{1}{2} \, \int d^4 x \, (\partial \varphi)^2
\;,
\end{equation}
while $\delta_L$ ($\delta_R$) imposes Dirichlet boundary conditions on the
$L$ ($R$) surface. Because of several reasons, it becomes convenient to
generalize, at this point, the boundary condition on the
$R$ mirror, by assuming that $\eta$ may also depend on $x_0$, the Euclidean
time. Namely, \mbox{$\eta({\mathbf x_\parallel}) \to \eta(x_\parallel)$}.
One of the reasons for doing this is that the results thus obtained
become more symmetrical, in the sense that the coordinates
$x_\parallel$ shall appear on an equal footing.  Besides, as a byproduct,
one could make use of those results in dynamical Casimir effect situations.
Note that, strictly speaking, ${\mathcal Z}$ will not be a quantum
partition function when $\eta$ depends on time. We do keep, however, the
same notation (${\mathcal Z}$) for the resulting object. 
Static Casimir effect results will be obtained by simply dropping the time
dependence of $\eta$, at the end of the calculation. 

Finally, note that ${\mathcal Z}$, evaluated with a time dependent $\eta$,
may also be interpreted as the classical partition function for a {\em
static\/} system in $4$ spatial dimensions, in the presence of two (static) boundaries.  

To proceed, note that the two functional delta functions may be exponentiated by
introducing two auxiliary fields, $\lambda_L$ and $\lambda_R$, to obtain
for ${\mathcal Z}$ the equivalent expression:
\begin{equation}
{\mathcal Z}\;=\; \int {\mathcal D}\varphi {\mathcal D}\lambda_L
\, {\mathcal D}\lambda_R \; e^{-S(\varphi;\lambda_L,\lambda_R)} ,
\end{equation}
with
\begin{eqnarray}
	&&S(\varphi; \lambda_L,\lambda_R) = S_0(\varphi) -
				 i \int d^4x \big[ \lambda_L(x_\parallel) \delta(x_3)
				 \nonumber\\ &&
+ \lambda_R(x_\parallel) \delta(x_3-\psi(x_\parallel)) \big] \varphi(x)\; ,
\end{eqnarray}
where \mbox{$\psi(x_\parallel) \equiv a + \eta(x_\parallel)$}. 

Integrating out $\varphi$, we see that a ${\mathcal Z}_0$ factor,
the partition function for the field $\varphi$ in the absence of boundary
conditions, factors out, while the rest is an integral over the auxiliary fields:
\begin{eqnarray}
&&{\mathcal Z}\;=\; {\mathcal Z}_0 
\int {\mathcal D}\lambda_L \, {\mathcal D}\lambda_R
\nonumber\\ && 
e^{-\frac{1}{2} \int d^3x_\parallel \int d^3y_\parallel 
\sum_{\alpha,\beta}\lambda_\alpha(x_\parallel)  
{\mathbb T}_{\alpha\beta}(x_\parallel,y_\parallel) 
\lambda_\beta(y_\parallel)}\; ,
\end{eqnarray}
where $\alpha,\,\beta=L,R$ and we have introduced the object:
\begin{align}
T_{LL}(x_\parallel,y_\parallel) &=\langle
x_\parallel,0|(-\partial^2)^{-1} |y_\parallel,0\rangle \\
T_{LR}(x_\parallel,y_\parallel) &=\langle
x_\parallel,0|(-\partial^2)^{-1}
|y_\parallel,\psi(y_\parallel)\rangle \\
T_{RL}(x_\parallel,y_\parallel) &=\langle
x_\parallel,\psi(x_\parallel)|(-\partial^2)^{-1} 
|y_\parallel,0\rangle \\
T_{RR}(x_\parallel,y_\parallel) &=\langle
x_\parallel,\psi(x_\parallel)|(-\partial^2)^{-1} 
|y_\parallel,\psi(y_\parallel) \;.
\end{align}
We have used a `bra-ket' notation as a convenient way to denote
matrix elements of operators. $\partial^2$ is the four-dimensional
Laplacian.  Thus,
\begin{equation}
	\langle x|(-\partial^2)^{-1} |y\rangle \,=\, \int
	\frac{d^4k}{(2\pi)^4} \, \frac{e^{i k \cdot (x-y)}}{k^2}\;.
\end{equation}
The four kernels $T_{\alpha \beta}$ can be naturally regarded as matrix
elements of a $2\times 2$ kernel matrix: 
\mbox{${\mathbb T} \equiv \big(T_{\alpha\beta}\big)_{\alpha,\beta=L,R}$}. 

The vacuum energy of the system, $E_{\rm vac}$, obtained by subtracting the
zero-point energy of a free field with trivial boundary conditions at
infinity, may then be obtained as follows:
\begin{equation}\label{eq:evac}
	E_{\rm vac}\;=\; \lim_{T \to \infty} 
	\big[ \frac{\Gamma(\eta)}{T} \big]\Big|_{\eta = \eta({\mathbf
	x_\parallel})} \;,
\end{equation}
where $T$ is the extent of the time dimension, and $\Gamma$ is given by:
\begin{equation}
	\Gamma \equiv - \log\big(\frac{\mathcal Z}{{\mathcal Z}_0}\big) = 
\frac{1}{2} {\rm Tr} \big(\log {\mathbb T}\big) \;,
\end{equation}
wich resembles an effective action in the presence of a background field
$\eta(x_\parallel)$.
Note that the trace above is meant to act on both discrete and continuous indices.

The Casimir energy is then obtained from the vacuum energy $E_{\rm vac}$, by
simply discarding contributions due to the vacuum field distortion produced when 
the mirrors are infinitely far apart. This `self-energy' piece, is irrelevant
for the kind of physical observable we have in mind here, and shall
therefore be subtracted. 

In the next section, we construct the expansion of $\Gamma$ in powers of
$\eta$, whence the information about the Casimir energy will be extracted
from (\ref{eq:evac}), after discarding self-energy terms.
\section{Perturbative expansion}\label{sec:expansion} 
The procedure to obtain the formal expression for the expansion of $\Gamma$
is rather straightforward. Indeed, assuming that ${\mathbb T}^{(i)}$
denotes the term of order $i$ in the expansion of ${\mathbb T}$ in powers of $\eta$,
\begin{equation}
{\mathbb T}=
{\mathbb T}^{(0)}+{\mathbb T}^{(1)}+{\mathbb T}^{(2)}+
{\mathbb T}^{(3)}+{\mathbb T}^{(4)}+\ldots\, ,
\end{equation}
we obtain an expansion for $\Gamma$ of the form
\begin{equation}
\Gamma\,=\,\Gamma^{(0)}+\Gamma^{(1)}+\Gamma^{(2)}+\Gamma^{(3)}+\Gamma^{(4)}+\ldots 
\end{equation}
Terms of up to the second order for $\Gamma$ are then given by 
\begin{equation}\label{eq:uptosecond} 
	\Gamma^{(0)}=\frac{1}{2} {\rm Tr}  \log \big({\mathbb
T}^{(0)}\big)\;,\;\; 
\Gamma^{(1)}=\frac{1}{2} {\rm Tr}\left[({\mathbb T}^{(0)})^{-1}
{\mathbb T^{(1)}}\right] 
\end{equation}
and
\begin{equation}
\Gamma^{(2)}= \Gamma^{(2,1)} \,+\, \Gamma^{(2,2)} 
\end{equation}
where:
\begin{eqnarray}
\Gamma^{(2,1)}&=&\frac{1}{2} {\rm Tr}\left[({\mathbb T}^{(0)})^{-1}{\mathbb T}^{(2)}\right]
\nonumber\\
\Gamma^{(2,2)}&=& -\frac{1}{4} {\rm Tr}\left[ ({\mathbb T}^{(0)})^{-1}{\mathbb T}^{(1)}  ({\mathbb T}^{(0)})^{-1}{\mathbb T}^{(1)} \right] \;.
\end{eqnarray}

Collecting all the terms of third order in $\eta$, we see that
$\Gamma^{(3)}$ may be expressed as the sum of three contributions:
\begin{equation}
	\Gamma^{(3)} \,=\, \Gamma^{(3,1)} \,+\, \Gamma^{(3,2)} \,+\,
	\Gamma^{(3,3)} \;, 
\end{equation}
where
\begin{equation}\label{eq:defg31}
	\Gamma^{(3,1)}= \frac{1}{2} {\rm Tr}\left[({\mathbb
	T}^{(0)})^{-1}{\mathbb T}^{(3)}\right] \;,
\end{equation}
\begin{equation}\label{eq:defg32}
	\Gamma^{(3,2)} = - \frac{1}{2} {\rm Tr}\left[ ({\mathbb T}^{(0)})^{-1}{\mathbb T}^{(1)}  ({\mathbb
T}^{(0)})^{-1}{\mathbb T}^{(2)} \right] \;, 
\end{equation}
and
\begin{equation}\label{eq:defg33}
	\Gamma^{(3,3)}= \frac{1}{6} {\rm Tr}\left[ ({\mathbb T}^{(0)})^{-1}{\mathbb T}^{(1)}  
({\mathbb T}^{(0)})^{-1}{\mathbb T}^{(1)} ({\mathbb T}^{(0)})^{-1}{\mathbb T}^{(1)}\right] \;.
\end{equation}

Finally, $\Gamma^{(4)}$ is the sum of five independent terms, which in turn
result from collecting all the terms of fourth order in the expansion for
$\Gamma$:
\begin{equation}\label{eq:gfourdeco}
	\Gamma^{(4)} \;=\; \sum_{l=1}^5 \Gamma^{(4,l)} \;,
\end{equation}
with:
\begin{equation}\label{eq:defg41}
	\Gamma^{(4,1)}= \frac{1}{2} {\rm Tr}\left[({\mathbb
	T}^{(0)})^{-1}{\mathbb T}^{(4)}\right] \;,
\end{equation}
\begin{equation}\label{eq:defg42}
	\Gamma^{(4,2)} = - \frac{1}{2} {\rm Tr}\left[ ({\mathbb T}^{(0)})^{-1}{\mathbb T}^{(1)}  ({\mathbb
T}^{(0)})^{-1}{\mathbb T}^{(3)} \right] \;, 
\end{equation}
\begin{equation}\label{eq:defg43}
	\Gamma^{(4,3)} = - \frac{1}{4} {\rm Tr}\left[ ({\mathbb T}^{(0)})^{-1}{\mathbb T}^{(2)}  ({\mathbb
T}^{(0)})^{-1}{\mathbb T}^{(2)} \right] \;, 
\end{equation}
\begin{equation}\label{eq:defg44}
	\Gamma^{(4,4)}= \frac{1}{2} {\rm Tr}\left[ ({\mathbb T}^{(0)})^{-1}{\mathbb T}^{(1)}  
({\mathbb T}^{(0)})^{-1}{\mathbb T}^{(1)} ({\mathbb T}^{(0)})^{-1}{\mathbb
T}^{(2)}\right] \;,
\end{equation}
and
\begin{eqnarray}\label{eq:defg45}
	&&\Gamma^{(4,5)}= -\frac{1}{8} \times \\ && {\rm Tr}\left[ ({\mathbb T}^{(0)})^{-1}{\mathbb T}^{(1)}  
({\mathbb T}^{(0)})^{-1}{\mathbb T}^{(1)} ({\mathbb T}^{(0)})^{-1}{\mathbb T}^{(1)}
({\mathbb T}^{(0)})^{-1}{\mathbb T}^{(1)}\right] \nonumber .
\end{eqnarray}

To proceed to the evaluation of the explicit form for each term,
it is convenient to introduce ${\tilde \eta}(k_\parallel)$, the Fourier
transform of $\eta(x_\parallel)$,
\begin{equation}\label{eq:defftranseta}
	{\tilde \eta}(k_\parallel)\,=\, \int d^3x_\parallel \, 
	e^{-i k_\parallel \cdot x_\parallel} \, \eta(x_\parallel) \;.
\end{equation}
Also, we note that $({\mathbb T}^{(0)})^{-1}$, which is translation
invariant, may be written as follows:
\begin{eqnarray}
	({\mathbb T}^{(0)})^{-1}(x_\parallel,x'_\parallel) &=& 
	({\mathbb T}^{(0)})^{-1}(x_\parallel-x'_\parallel)\nonumber \\ &=& 
	\int \frac{d^3k_\parallel}{(2\pi)^3} \, e^{i k_\parallel \cdot
	(x_\parallel-x'_\parallel)} \,
	\widetilde{\mathbb D}(k_\parallel) \;,
\end{eqnarray}
with:
\begin{equation}
	\widetilde{\mathbb D}(k_\parallel) =
	\frac{2 |k_\parallel|}{1- e^{-2 |k_\parallel| a}} \,
	\left( 
		\begin{array}{cc}
		1 & - e^{-|k_\parallel| a} \\
		- e^{-|k_\parallel| a} & 1
		\end{array}
	\right) \;.
\end{equation}
Another important ingredient is the form of the matrix elements of
${\mathbb T}$ to the desired order. Regarding the diagonal elements, one
sees that $T_{LL}^{(j)}=0$ for $j>0$, while $T_{RR}^{(j)}=0$ for odd $j$.
Thus, for the diagonal elements, we need:
\begin{equation}
	T_{RR}^{(2)}(x_\parallel, x'_\parallel) = 
	\int \frac{d^3k_\parallel}{(2\pi)^3} e^{i k_\parallel
	\cdot (x_\parallel-x'_\parallel)} \, \frac{|k_\parallel|}{4} \,
	\big[\eta(x_\parallel) - \eta(x'_\parallel) \big]^2
\end{equation}
and 
\begin{equation}
	T_{RR}^{(4)}(x_\parallel, x'_\parallel) = 
	\int \frac{d^3k_\parallel}{(2\pi)^3} e^{i k_\parallel
	\cdot (x_\parallel-x'_\parallel)} \, \frac{|k_\parallel|^3}{48} \,
	\big[\eta(x_\parallel) - \eta(x'_\parallel) \big]^4 \;.
\end{equation}
For the off-diagonal matrix elements, their general form at an arbitrary
order may be written in a rather compact form. Indeed,
\begin{eqnarray}
	T_{LR}^{(j)}(x_\parallel, x'_\parallel) = \frac{(-1)^j}{2 j!} \,
	&\int \frac{d^3k_\parallel}{(2\pi)^3} &e^{i k_\parallel \cdot (x_\parallel-x'_\parallel)} 
	\, e^{- |k_\parallel| a} \, |k_\parallel|^{j-1} \, \nonumber \\
	&\times&\big[\eta(x'_\parallel)\big]^j \;,
\end{eqnarray}
\begin{equation}
	T_{RL}^{(j)}(x_\parallel, x'_\parallel) =
	T_{LR}^{(j)}(x'_\parallel, x_\parallel) \;. 
\end{equation}

\section{Results}\label{sec:results}
Although the terms of order zero, one and two are already known, we
present, for the sake of completeness, the results corresponding to those
orders as well.
\subsection{Zeroth order}\label{ssec:zero}
The zeroth-order term is simply obtained by recalling the form of
$\Gamma^{(0)}$, from which one just needs to subtract its $a \to \infty$ limit.
After that subtraction, one obtains a result that may be written in the following way:
\begin{equation}
	\lim_{T\to\infty} \Big[\frac{\Gamma^{(0)}}{T }\Big] \;=\;
	\frac{L^2}{2} \,\int\frac{d^3p_\parallel}{(2\pi)^3}
	\log\big(1-e^{-2 p_\parallel a}\big)
\end{equation}
where $L$ is a length that measures the size of the mirror along the
parallel directions.

Thus, the Casimir energy due to this term, $E_{\rm vac}^{(0)}$, 
becomes:
\begin{equation}
	E_{\rm vac}^{(0)}\,=\,\frac{L^2}{2} \,\int\frac{d^3p_\parallel}{(2\pi)^3}
	\log\big(1-e^{-2 |p_\parallel|a }\big)=-\frac{\pi^2 L^2}{1440 \, a^3}\, ,
\end{equation}
as expected.
\subsection{First order}\label{ssec:first}
It is quite straightforward to compute the first order term $\Gamma^{(1)}$. 
Introducing a function $B$, defined as:
\begin{equation}
	B(q_\parallel) \,=\, \frac{1}{e^{2 |q_\parallel| a} -1 } \;,
\end{equation}
where $q_\parallel$ is a $3$-vector, we obtain 
\begin{eqnarray}
	\Gamma^{(1)} \,&=&\, \int \frac{d^3p_\parallel}{(2\pi)^3}
	\, B(p_\parallel) \, |p_\parallel| \, \times 
	\int d^3x_\parallel \, \eta(x_\parallel)\nonumber\\
	&=&\, \frac{\pi^2}{480 a^4}\int d^3x_\parallel \, \eta(x_\parallel)	
	 \; .
	\label{gamma1}
\end{eqnarray}

Since, by definition, $\eta$ measures the {\em departure\/} from a flat 
parallel mirrors configuration, one could impose on it the condition that the integral 
$\int d^3x_\parallel \, \eta(x_\parallel)$ equals zero. Were it not the
case, this condition could nevertheless have been achieved by subtracting a
constant (which has to be added to $a$) from $\eta$. 
Thus, in this case $\Gamma^{(1)} \,=\, 0$.
\subsection{Second order}\label{ssec:second}
The second order result has been known for quite some time. In Fourier
space, it may be written as follows:
\begin{equation}\label{eq:second}
	\Gamma^{(2)} \,=\, \frac{1}{2} \, \int
	\frac{d^3k_\parallel}{(2\pi)^3} \, f^{(2)}(k_\parallel)
	\big|{\widetilde\eta}(k_\parallel)\big|^2 \;, 
\end{equation}
with:
\begin{eqnarray}
	f^{(2)}(k_\parallel) &\,=\,& - 2 \int \frac{d^3p_\parallel}{(2\pi)^3}
	\, \big[ B(p_\parallel) B(p_\parallel + k_\parallel)  +  B(p_\parallel + k_\parallel) \big] \nonumber \\
&\times &	\, |p_\parallel|\, |p_\parallel + k_\parallel| \;.
	\label{f2}
\end{eqnarray}
From the structure of (\ref{eq:second}), it seems that, for the
perturbative expansion to make sense, an extra necessary condition (besides
$|\eta| << a$) is that the integral over $k_\parallel$ has to be
well-defined. 
However, we can, and will, consider cases where ${\widetilde\eta}(k_\parallel)$
is a generalized function. In particular,  $\delta$-like, concentrated
around a particular value of $k_\parallel$. In this case, all the correction terms
(not just the second order one) become proportional to the size of the
system, due to the periodicity of the perturbation.  Of course, the same is
always true of the zeroth order term. Thus, even for this singular case,
one could make sense of the perturbative corrections, at least if there is
a region where the corrections are reasonably small, in comparison with the
zeroth order term, after factorizing the (common) spatial size factor.
 
On the other hand, when ${\widetilde\eta}(k_\parallel)$ is a regular
function, a nontrivial condition emerging from (\ref{eq:second}) proceeds
from requiring its convergence at large momentum. It may be seen that, at
second order, this amounts to:
\begin{equation}
\int \frac{d^3k_\parallel}{(2\pi)^3} \, |k_\parallel|\,
	\big|{\widetilde\eta}(k_\parallel)\big|^2 \;<\; \infty \;.
\end{equation}
In a realistic situation, the Fourier spectrum of the deformation $\eta$
should have a cutoff; for instance,  a continuous description of the
mirrors should, at some point, be replaced by a discrete (lattice) one,
which introduces a large momentum cutoff of the order of the lattice
spacing. 

\subsection{Third order}
The contribution denoted by $\Gamma^{(3,1)}$ is {\em ultralocal}, understanding by
that an integral of a term involving (three) $\eta$'s at a single point
(without derivatives). The first example of an ultralocal contribution
was the first order term, but it could, as we have seen, be assumed to be
equal to zero. On the contrary, $\Gamma^{(3,1)}$, given explicitly by:
\begin{equation}
	\Gamma^{(3,1)} \,=\, \frac{1}{3!} \, \int \frac{d^3p_\parallel}{(2\pi)^3}
		B(p_\parallel) \, |p_\parallel|^3 \times 
	\int d^3x_\parallel \, \big[\eta(x_\parallel)\big]^3 
	\,\equiv \, \Gamma^{(3)}_l \label{Gamma31}
\end{equation}
does not necessarily vanish. We have used $\Gamma^{(3)}_l$, to denote
this contribution, the only ultralocal term at the third order. 

On the other hand, $\Gamma^{(3,2)}$ is `bilocal', i.e., it is cubic in
$\eta$ and involves products of $\eta$ at two (eventually equal) points. It
is the only bilocal piece of $\Gamma^{(3)}$. Its explicit form is
\begin{equation}\label{Gamma32}
\Gamma^{(3,2)} = \frac{1}{2} \int  \frac{d^3k_\parallel}{(2\pi)^3}
\, f^{(3)}_b(k_\parallel) \, {\widetilde\eta}(k_\parallel) \,
{\widetilde{\eta^2}}(-k_\parallel)  \,\equiv\, \Gamma^{(3)}_b \;,
\end{equation}
(note the Fourier transform of $\eta^2$), where we introduced $f^{(3)}_b$,
the bilocal kernel at the third order:
\begin{equation}
f^{(3)}_b(k_\parallel) \,=\,- \int  \frac{d^3p_\parallel}{(2\pi)^3} \,
B(p_\parallel) \; |p_\parallel| \,
|p_\parallel+k_\parallel|^2 \;.
\label{f3b}
\end{equation}

Finally, $\Gamma^{(3,3)}$ is trilocal, since the three $\eta$'s appear at
(generally) different spacetime points:
\begin{eqnarray}\label{Gamma33}
	\Gamma^{(3,3)} &=& \frac{1}{3!} 
	\int  \frac{d^3k_\parallel}{(2\pi)^3} \frac{d^3q_\parallel}{(2\pi)^3}
\, f^{(3)}_t(k_\parallel,q_\parallel) \, 
{\widetilde\eta}(k_\parallel) \, {\widetilde{\eta}}(q_\parallel)  \nonumber \\
&\times & {\widetilde{\eta}}(-k_\parallel - q_\parallel)  
\equiv \Gamma^{(3)}_t \;,
\end{eqnarray}
where the $f^{(3)}$ kernel is given by:
\begin{eqnarray}
&&f^{(3)}_t(k_\parallel,l_\parallel)  = 2 \, \int \frac{d^3p_\parallel}{(2\pi)^3} 
|k_\parallel + l_\parallel + p_\parallel|\,
|l_\parallel + p_\parallel|\,
|p_\parallel| \nonumber \\
& \times &  
\Big[ 4 B(k_\parallel + l_\parallel + p_\parallel) B(l_\parallel +
p_\parallel) B(p_\parallel) \nonumber \\ &+& B(k_\parallel + l_\parallel + p_\parallel)
	B(l_\parallel + p_\parallel) \\ &+& 5 B(k_\parallel + l_\parallel + p_\parallel) B(p_\parallel)+
	3 B(k_\parallel + l_\parallel 
+ p_\parallel) \Big] \; \nonumber.
\label{f3t}
\end{eqnarray}

The full third order term is then given by the sum of the previously
derived local, bilocal and trilocal contributions:
\begin{equation}\label{eq:resO3}
	\Gamma^{(3)} \,=\, \Gamma^{(3)}_{l} \,+\, \Gamma^{(3)}_{b}\,+\, 
\Gamma^{(3)}_{t}\;. 
\end{equation}
\subsection{Fourth order}\label{ssec:fourth}
As we have just done for the third order contribution, we present the
partial results corresponding to what we denoted as $\Gamma^{(4,l)}$, for
the different values of $l$, which run from $1$ to $5$.  

The $\Gamma^{(4,1)}$ term, given by (\ref{eq:defg41}), contains both local
and bilocal parts, the local one being $a$-independent. After discarding the
$a$-independent part, the explicit form of the term becomes:
\begin{eqnarray}\label{eq:g41res}
\Gamma^{(4,1)} &=&\frac{1}{2} \, \int \frac{d^3k_\parallel}{(2\pi)^3}
f^{(4,1)}(k_\parallel) \,
\big[{\widetilde{\eta^2}}(k_\parallel)  {\widetilde{\eta^2}}(-k_\parallel) \nonumber \\
&-& \frac{4}{3} {\widetilde{\eta^3}}(-k_\parallel)
{\widetilde{\eta}}(k_\parallel) \big]
\end{eqnarray}
with 
\begin{equation}\label{eq:g41ker}
	f^{(4,1)}(k_\parallel) \, =\, \frac{1}{4} \, 
	\int \frac{d^3p_\parallel}{(2\pi)^3} \, B(p_\parallel)
\, |p_\parallel| \,|p_\parallel + k_\parallel|^3 \;.
\end{equation}

The next term, $\Gamma^{(4,2)}$ may be written as follows:
\begin{equation}\label{eq:g42res}
\Gamma^{(4,2)} =\frac{1}{2} \, \int \frac{d^3k_\parallel}{(2\pi)^3}
f^{(4,2)}(k_\parallel) \, {\widetilde{\eta^3}}(-k_\parallel)
{\widetilde{\eta}}(k_\parallel) \;.
\end{equation}
with
\begin{eqnarray}\label{eq:resg42}
f^{(4,2)}(k_\parallel) &=& -\frac{4}{3} \, f^{(4,1)}(k_\parallel) \\
	&-&\frac{2}{3} \, \int \frac{d^3p_\parallel}{(2\pi)^3} \, 
B(p_\parallel+k_\parallel) B(p_\parallel) \;
\, |p_\parallel| |p_\parallel + k_\parallel|^3 \;.\nonumber 
\end{eqnarray}

We now write the result corresponding to $\Gamma^{(4,3)}$, which involves
three different kernels:
\begin{eqnarray}
	\Gamma^{(4,3)} &=& \frac{1}{4!} \,
	\int \frac{d^3k_\parallel}{(2\pi)^3} \frac{d^3q_\parallel}{(2\pi)^3} \frac{d^3l_\parallel}{(2\pi)^3}
\,f_1^{(4,3)}(k_\parallel,q_\parallel,l_\parallel) \,
{\widetilde{\eta}}(k_\parallel) \nonumber \\ &\times &{\widetilde{\eta}}(q_\parallel) {\widetilde{\eta}}(l_\parallel) 
{\widetilde{\eta}}(-k_\parallel-q_\parallel-l_\parallel) \nonumber\\
	&+& \frac{1}{3!} \,
	\int \frac{d^3k_\parallel}{(2\pi)^3} \frac{d^3q_\parallel}{(2\pi)^3}
\,f_2^{(4,3)}(k_\parallel,q_\parallel) \,
{\widetilde{\eta}}(k_\parallel) {\widetilde{\eta}}(q_\parallel)\nonumber \\ &\times &
{\widetilde{\eta^2}}(-k_\parallel-q_\parallel) \nonumber\\
	&+& \frac{1}{2!} \,
	\int \frac{d^3k_\parallel}{(2\pi)^3} 
	\,f_3^{(4,3)}(k_\parallel) \, {\widetilde{\eta^2}}(k_\parallel)
	{\widetilde{\eta^2}}(-k_\parallel) \;.
\end{eqnarray}
After some rather lengthy algebraic manipulations, we see that the form of
the kernels introduced above is the following (again, discarding $a$-independent terms):
$$
f_1^{(4,3)}(k_\parallel,q_\parallel,l_\parallel) = - 6 \int \frac{d^3p_\parallel}{(2\pi)^3} 
\big[
B(k_\parallel + q_\parallel + l_\parallel + p_\parallel) B(l_\parallel + p_\parallel) 
$$
$$
+ B(k_\parallel + q_\parallel + l_\parallel + p_\parallel) 
+ B(l_\parallel + p_\parallel)\big]
$$
\begin{equation}
\times | k_\parallel + q_\parallel + l_\parallel + p_\parallel |
|q_\parallel + l_\parallel + p_\parallel | | l_\parallel + p_\parallel |
|p_\parallel |\;,
\end{equation} 
\begin{equation}
	f_2^{(4,3)}(k_\parallel,q_\parallel) = 3 \int \frac{d^3p_\parallel}{(2\pi)^3} \,
B(p_\parallel) \, |k_\parallel + q_\parallel + p_\parallel |^2 |q_\parallel + p_\parallel | |p_\parallel | 
\end{equation} 
and
\begin{equation}
	f_3^{(4,3)}(k_\parallel) \;= \; - f^{(4,1)}(k_\parallel) \;.
\end{equation} 
		
Let us present now the results corresponding to the $\Gamma^{(4,4)}$
contribution. We have found that it contains both $4$-local and $3$-local terms:
\begin{eqnarray}
	\Gamma^{(4,4)} &=& \frac{1}{4!} \,
	\int \frac{d^3k_\parallel}{(2\pi)^3} \frac{d^3q_\parallel}{(2\pi)^3} \frac{d^3l_\parallel}{(2\pi)^3}
\,f_1^{(4,4)}(k_\parallel,q_\parallel,l_\parallel) \,
{\widetilde{\eta}}(k_\parallel)\nonumber \\ &\times & {\widetilde{\eta}}(q_\parallel) {\widetilde{\eta}}(l_\parallel) 
{\widetilde{\eta}}(-k_\parallel-q_\parallel-l_\parallel) \nonumber\\
	&+& \frac{1}{3!} \,
	\int \frac{d^3k_\parallel}{(2\pi)^3} \frac{d^3q_\parallel}{(2\pi)^3}
\,f_2^{(4,4)}(k_\parallel,q_\parallel) \,
{\widetilde{\eta}}(k_\parallel) {\widetilde{\eta}}(q_\parallel)\nonumber \\ &\times &
{\widetilde{\eta^2}}(-k_\parallel-q_\parallel) \;,
\end{eqnarray}
with the kernels:
\begin{eqnarray}
&&	f_1^{(4,4)}(k_\parallel,q_\parallel,l_\parallel) =- 12 \, \int \frac{d^3p_\parallel}{(2\pi)^3} 
	| k_\parallel + q_\parallel + l_\parallel + p_\parallel |\nonumber \\ &\times &
	|q_\parallel + l_\parallel + p_\parallel |
	| l_\parallel + p_\parallel | |p_\parallel | \nonumber\\
	&\times& \Big[4 B(k_\parallel + q_\parallel + l_\parallel +
	p_\parallel) 
B(q_\parallel + l_\parallel + p_\parallel) B(l_\parallel + p_\parallel)
\nonumber\\ 
&+& 3 B(k_\parallel + q_\parallel + l_\parallel + p_\parallel)
B(q_\parallel + l_\parallel + p_\parallel)\nonumber\\
&+&   B(q_\parallel + l_\parallel + p_\parallel) B(l_\parallel +
p_\parallel) \\
&+& B(k_\parallel + q_\parallel + l_\parallel + p_\parallel) B(l_\parallel + p_\parallel) +
B(q_\parallel + l_\parallel + p_\parallel) \Big]\;,\nonumber 
\end{eqnarray}
and
\begin{eqnarray}
	f_2^{(4,4)}(k_\parallel,q_\parallel) &=& 3\, \int \frac{d^3p_\parallel}{(2\pi)^3} 
	| k_\parallel + q_\parallel + p_\parallel |
	|q_\parallel + p_\parallel |
	|p_\parallel|^2  \nonumber\\
	&\times& \Big[3 B(k_\parallel + q_\parallel + p_\parallel) 
B(q_\parallel + p_\parallel) \nonumber \\ &-& B(q_\parallel + p_\parallel) B(p_\parallel)
+ B(q_\parallel + p_\parallel) \Big]\;.
\end{eqnarray}
The $\Gamma^{(4,5)}$ term is $4$-local:
\begin{eqnarray}
	\Gamma^{(4,5)} &=& \frac{1}{4!} \,
	\int \frac{d^3k_\parallel}{(2\pi)^3} \frac{d^3q_\parallel}{(2\pi)^3} \frac{d^3l_\parallel}{(2\pi)^3}
\,f^{(4,5)}(k_\parallel,q_\parallel,l_\parallel) \,
{\widetilde{\eta}}(k_\parallel)\nonumber \\ &\times & {\widetilde{\eta}}(q_\parallel) {\widetilde{\eta}}(l_\parallel) 
{\widetilde{\eta}}(-k_\parallel-q_\parallel-l_\parallel) \;,
\end{eqnarray}
with:
\begin{eqnarray}
&&f^{(4,5)}(k_\parallel,q_\parallel,l_\parallel) = - 6 \, \int \frac{d^3p_\parallel}{(2\pi)^3} 
| k_\parallel + q_\parallel + l_\parallel + p_\parallel |
\nonumber \\  &\times & |q_\parallel + l_\parallel + p_\parallel |
| l_\parallel + p_\parallel | |p_\parallel | \nonumber \\
&\times & \Big[ 8  
B(k_\parallel + q_\parallel + l_\parallel + p_\parallel)
B(q_\parallel + l_\parallel + p_\parallel)
B(l_\parallel + p_\parallel) 
B(p_\parallel)
\nonumber\\
&+& 5\, B(k_\parallel + q_\parallel + l_\parallel + p_\parallel)
B(q_\parallel + l_\parallel + p_\parallel)
B(l_\parallel + p_\parallel) 
\nonumber\\
&+& 2 \, B(k_\parallel + q_\parallel + l_\parallel + p_\parallel)
B(q_\parallel + l_\parallel + p_\parallel)
B(p_\parallel)
\nonumber\\
&+& B(k_\parallel + q_\parallel + l_\parallel + p_\parallel)
B(l_\parallel + p_\parallel) B(p_\parallel) \nonumber\\
&+& B(k_\parallel + q_\parallel + l_\parallel + p_\parallel)
B(l_\parallel + p_\parallel) 
\Big]\;.
\end{eqnarray}
\section{Low and high momentum expansions: connection with the  PFA}\label{sec:other}
In order to discuss the forthcoming approximations, we find it convenient
to first write the previously considered perturbative expansion as follows:
\begin{eqnarray}
E_{\rm vac}&=&-\frac{\pi^2 L^2}{1440 a^3}+\frac{1}{T}\sum_{n\geq 1}\frac{1}{a^{3+n}}\int \frac{d^3k_\parallel^{(1)}}{(2\pi)^3}
...\frac{d^3k_\parallel^{(n)}}{(2\pi)^3}\nonumber \\ &\times &
\delta(k_\parallel^{(1)}+...+k_\parallel^{(n)}) h^{(n)}(ak_\parallel^{(1)},...,ak_\parallel^{(n)})\nonumber\\
&\times& {\widetilde{\eta}}(k^{(1)}_\parallel) ...{\widetilde{\eta}}(k^{(n)}_\parallel)\, .
\label{genexp}
\end{eqnarray}
The explicit form of the form factors $h^{(n)}$, up to $n=4$, is
completely determined by the results obtained in the previous
section. Indeed, one just needs to include the proper $\delta$ function
factors in some of the terms we calculated, before adding them.  In the equation above we have made explicit
the fact that the (dimensionless) functions $h^{(n)}$ depend on the dimensionless variables $ak_\parallel^{(j)},\; j=1,...,n$.

Note that (\ref{genexp}) has the structure of the general Taylor expansion of a
functional in terms of its argument, an expression entirely analogous to
the one we could use for the expansion of the effective action in a $2+1$
dimensional quantum field theory, in terms of the `field' $\eta$, the form
factors $h^{(n)}$ playing the role of the $n$-point vertex functions.

\subsection{PFA and the low momentum expansion}

Let us assume here that the function $\eta({\mathbf x_\parallel})$ is
slowly varying. In terms of its Fourier transform, it will be peaked at
zero momentum; therefore we can approximate
$h^{(n)}(ak_\parallel^{(1)},...,ak_\parallel^{(n)})\simeq h^{(n)}(0,...,0)$. 
As a consequence:
\begin{equation}
E_{\rm vac}\simeq-\frac{\pi^2 L^2}{1440 a^3}+\sum_{n\geq 1}\frac{1}{a^{3+n}}h^{(n)}(0,..,0)\int d^2{\mathbf x_\parallel} \eta({\mathbf x_\parallel})^n
\label{low}
\end{equation}
In principle, one could evaluate the first terms by using the
results of Sec.\ref{sec:results}. However, there is also a shortcut: for a
constant $\eta({\mathbf x_\parallel})=\eta_0$, the vacuum energy is simply
given by Eq.(\ref{low}) with the replacement $\int
d^2{\mathbf x_\parallel} \,\eta({\mathbf x_\parallel})^n\rightarrow
L^2\eta_0^n$. 
But for this particular case we know, of course, that the answer is the
Casimir energy between parallel plates separated by a distance $a+\eta_0$.
Therefore, in the low momentum approximation the perturbative series can be
summed up, the result being 
\begin{equation}
E_{\rm vac}\simeq - \frac{\pi^2}{1440}\int 
\frac{d^2{\mathbf x_\parallel}}{\left(a+\eta({\mathbf x_\parallel})\right)^3}\, ,
\end{equation} 
which agrees with the PFA. 

Note also that the above suggests a nontrivial consistency check for our
calculations: indeed, one can compute the first terms of the series in
Eq.(\ref{low}) using the explicit expressions given in Sec.\ref{sec:results} for the 
different contributions to the effective action $\Gamma^{(i,j)}$.
We present some details of that calculation in the Appendix. 
The result is
$$
E_{\rm vac}\simeq -\frac{\pi^2}{1440 a^3} \, \int 
d^2{\mathbf x_\parallel}
 \Big( 1-3\frac{\eta}{a}+ 6(\frac{\eta}{a})^2 
$$
\begin{equation}\label{cons}
-10(\frac{\eta}{a})^3 + 15(\frac{\eta}{a})^4\Big), 
\end{equation}
that agrees with the expansion in powers of $\eta$ of the PFA result.

One could also go beyond the PFA, by expanding the form factors around $k_\parallel
=0$.  In this way, one should be able to recover the derivative expansion
for the Casimir energy that we introduced in Ref.\cite{Fosco:2011xx}. Once more, it
is useful to have in mind the analogy with the usual expansions of the
effective action in quantum field theory: the expansion in powers of the
field corresponds to the perturbative expansion in powers of $\eta$, while
the derivative expansion  corresponds to an expansion in derivatives of
$\psi=a+\eta$, i.e. the improved PFA \cite{Fosco:2011xx}.  

\subsection{High momentum expansion}

We shall now consider the opposite limit, namely, $\vert k_\parallel\vert a\gg
1$, in which the scale of variation of the shape of the surface is much
shorter than the mean separation between surfaces. 

Let us begin by considering the second order term. In order to obtain the
large-momentum behavior of the form factor $f^{(2)}$ in Eq.(\ref{f2}), we
note that the first term is exponentially suppressed,
i.e., it can be written as $\vert k_\parallel\vert  e^{-2 a\vert k_\parallel\vert}$ 
times a convergent integral. After a shift in the integration variables, the second term can be approximated by 
\begin{eqnarray}
	f^{(2)}(k_\parallel) \, &\simeq & - 2 \int \frac{d^3p_\parallel}{(2\pi)^3}
	\,  B(p_\parallel)
	\, |p_\parallel|\, |p_\parallel - k_\parallel| \nonumber \\ &\simeq & - 2 \vert k_\parallel| \int \frac{d^3p_\parallel}{(2\pi)^3}
	\,  B(p_\parallel)
	\, |p_\parallel|\, \nonumber\\
	&=& -\frac{\pi^2}{240}\frac{\vert k_\parallel\vert}{a^4}\, .
	\label{f2ap}
\end{eqnarray}
Inserting this result into Eq.(\ref{eq:second}) we obtain
 \begin{equation}\label{eq:secondapprox}
	\Gamma^{(2)} \simeq\,- \frac{\pi^2}{480 a^4} \, \int
	\frac{d^3k_\parallel}{(2\pi)^3} \, \vert k_\parallel\vert
	\big|{\widetilde\eta}(k_\parallel)\big|^2 \;. 
\end{equation}
Unlike the low momentum expansion, the result is a nonlocal functional of
the shape of the surface. It can be rewritten in configuration space in
terms of the nonlocal operator $(-\nabla^2)^{1/2}$ as follows:
 \begin{equation}\label{eq:secondapproxconf}
	\frac{\Gamma^{(2)}}{T} \simeq\,- \frac{\pi^2 }{480 a^4} \, \int
	d^2{\mathbf x_\parallel}\, \eta({\mathbf x_\parallel}) (-\nabla^2)^{1/2}\eta({\mathbf x_\parallel})	\, .
	\end{equation}

We now consider the third order terms. From Eq.(\ref{f3b}) it is quite
straightforward to see that
\begin{equation}
f^{(3)}_b(k_\parallel) \,\simeq\,- \frac{\pi^2}{480}\frac{\vert k_\parallel\vert^2}{a^4}\, .
\end{equation}
Note that this contribution grows quadratically with the momentum and becomes local in configuration space:
 \begin{equation}
	\frac{\Gamma^{(3,2)}}{T} \simeq\, \frac{\pi^2}{960 a^4} \, \int d^2{\mathbf x_\parallel}\, \eta^2({\mathbf x_\parallel}) \nabla^2\eta({\mathbf x_\parallel})	 \;. 
\end{equation}
The trilocal contribution is determined by the form factor given in Eq.(\ref{f3t}). For high momenta, the leading contribution comes from the last term, wich is linear in $B$.
We obtain
\begin{equation}
f^{(3)}_t(k_\parallel,l_\parallel) \,\simeq\, \frac{\pi^2}{80}\frac{\vert k_\parallel (k_\parallel+l_\parallel)\vert}{a^4}\, ,
\end{equation}
which produces a nonlocal contribution to the vacuum energy
 \begin{eqnarray}
	\frac{\Gamma^{(3,3)}}{T} &\simeq & \frac{\pi^2 }{480 a^4} \, \int
	d^2{\mathbf x_\parallel}\, \eta({\mathbf x_\parallel}) (-\nabla^2)^{1/2}\eta({\mathbf x_\parallel})\nonumber \\ &\times & (-\nabla^2)^{1/2}\eta({\mathbf x_\parallel})	.
	\end{eqnarray} The  fourth order terms can be treated in a similar way. One can show that, to leading order,  they  grow cubically
with the momenta. 

\section{Examples}\label{sec:examples}
In this section we present some applications of the perturbative results obtained so far. As we will consider surfaces with periodic perturbations, the vacuum energy
will be proportional to the size of the system. Therefore, we will compute the vacuum energy per unit area ${\mathcal E_{\rm vac}}=E_{\rm vac}/L^2$.

\subsection{Sinusoidally corrugated surface}

We will first consider the simplest case of a sinusoidally corrugated surface $\eta( {\mathbf x_\parallel})=\epsilon \sin(q_1 x_1)$, with $\epsilon\ll a$.  For this particular corrugation,
the first and third order corrections do vanish, so on general grounds we expect
\begin{equation}\label{general}
{\mathcal E_{\rm vac}}\simeq -\frac{\pi^2 }{1440 a^3}\left
 [1 + g_2(q_1 a) \left(\frac{\epsilon}{a}\right)^2+ g_4(q_1 a) \left(\frac{\epsilon}{a}\right)^4+... \right]
\end{equation}
for some functions $g_2$ and $g_4$. The function $g_2$ has already been
calculated in Ref.\cite{perturb}. For the sake of completeness, we present
a plot of its numerical evaluation in Fig.1. In the zero momentum limit,
the numerical results reproduce the expected value, given in
Eq.(\ref{cons}). Moreover, the behavior for low momentum is quadratic
\cite{perturb,Fosco:2011xx}, which is consistent with the existence of a
derivative expansion for the Casimir energy \cite{Fosco:2011xx}. On the
other hand, in the limit $q_1 a\gg 1$, the plot reproduces the linear
behavior $g_2\sim q_1 a$ with the appropriate coefficient predicted in
Eq.(\ref{eq:secondapprox}).

\begin{figure}
\centering
\includegraphics[width=8cm , angle=0]{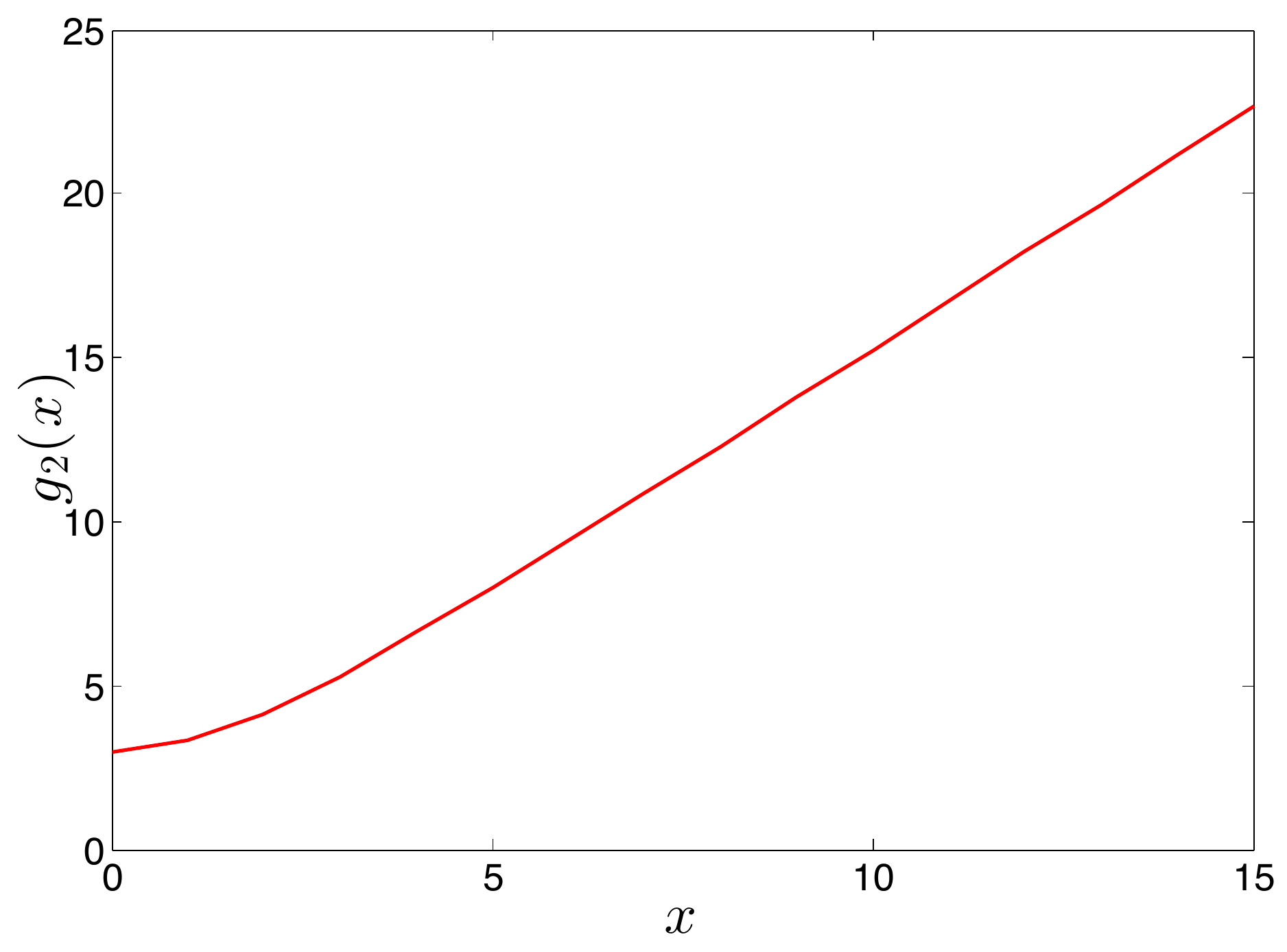}
\caption{\small{Plot of the function $g_2$ defined in Eq.(\ref{general}), as a function of the dimensionless variable $x=q_1 a$. Linear behavior of $g_2$ for large $x$ is shown.}} \label{fig1}
\end{figure}

The behavior of $g_4$ is shown in Fig.2. Once more, in the zero momentum
limit the numerical result is consistent with Eq.(\ref{cons}) (see inset in
Fig.2 for the small momentum behavior). On the other hand, in the high
momentum limit, $g_4\sim (q_1 a) ^3$. It is worth to note that the
perturbative expansion breaks down for very high momenta (short wavelength
of the corrugations). Indeed, the ratio of the fourth to second order
corrections is proportional to $(\epsilon q_1)^2$. Therefore,  the fourth
order correction becomes more relevant when approaching the short
wavelength limit. In Fig.3 we plot the ratio between the fourth and second
order correction coefficients, $g_4/g_2$, as a function of $x = q_1 a$.  
\begin{figure}
\centering
\includegraphics[width=8cm , angle=0]{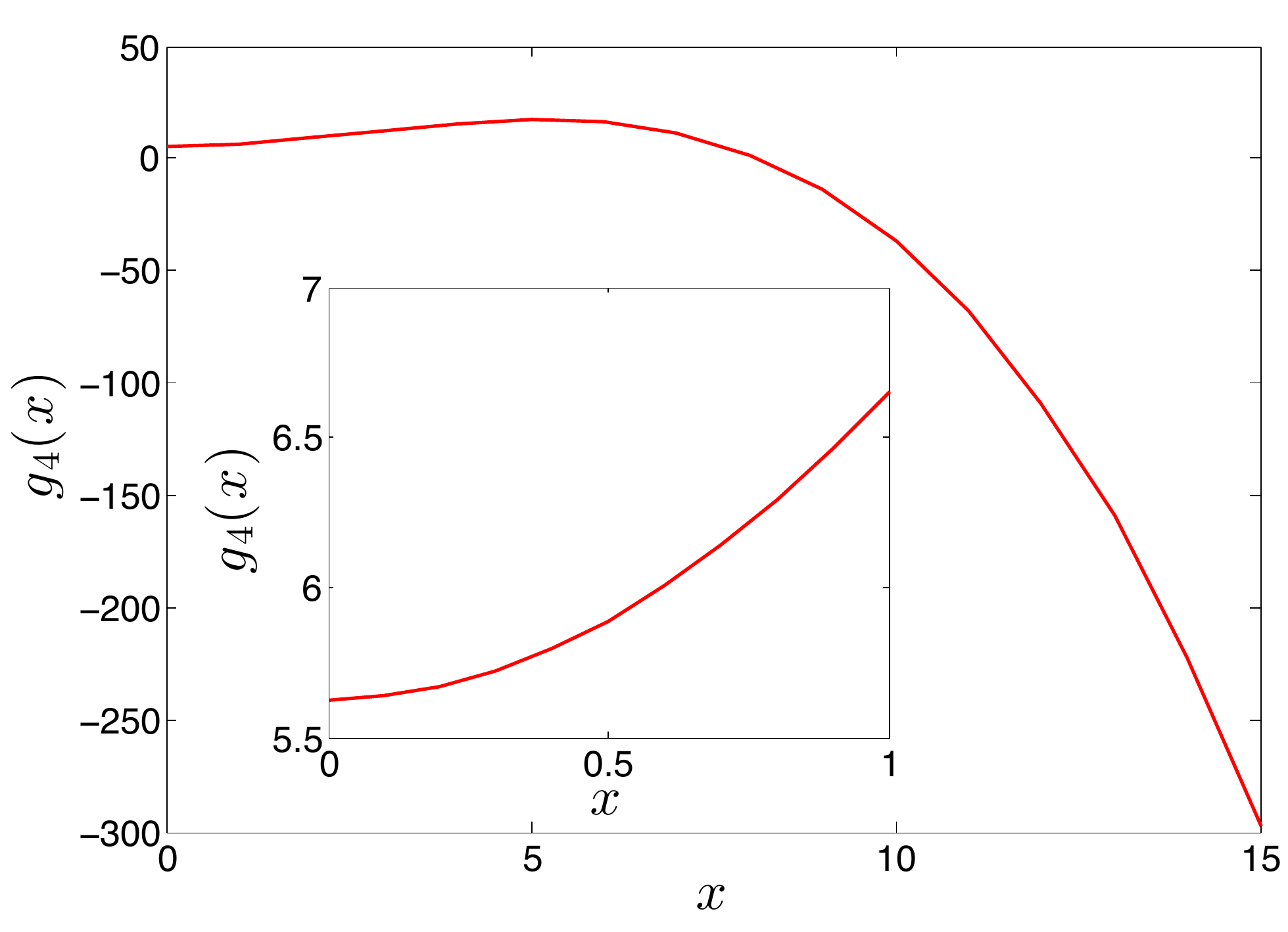}
\caption{\small{Plot of the function $g_4$ defined in Eq.(\ref{general}), as a function of the dimensionless variable $x=q_1 a$. The inset shows the low momentum behavior of $g_4$.}} \label{fig2}
\end{figure}

\begin{figure}
\centering
\includegraphics[width=8cm , angle=0]{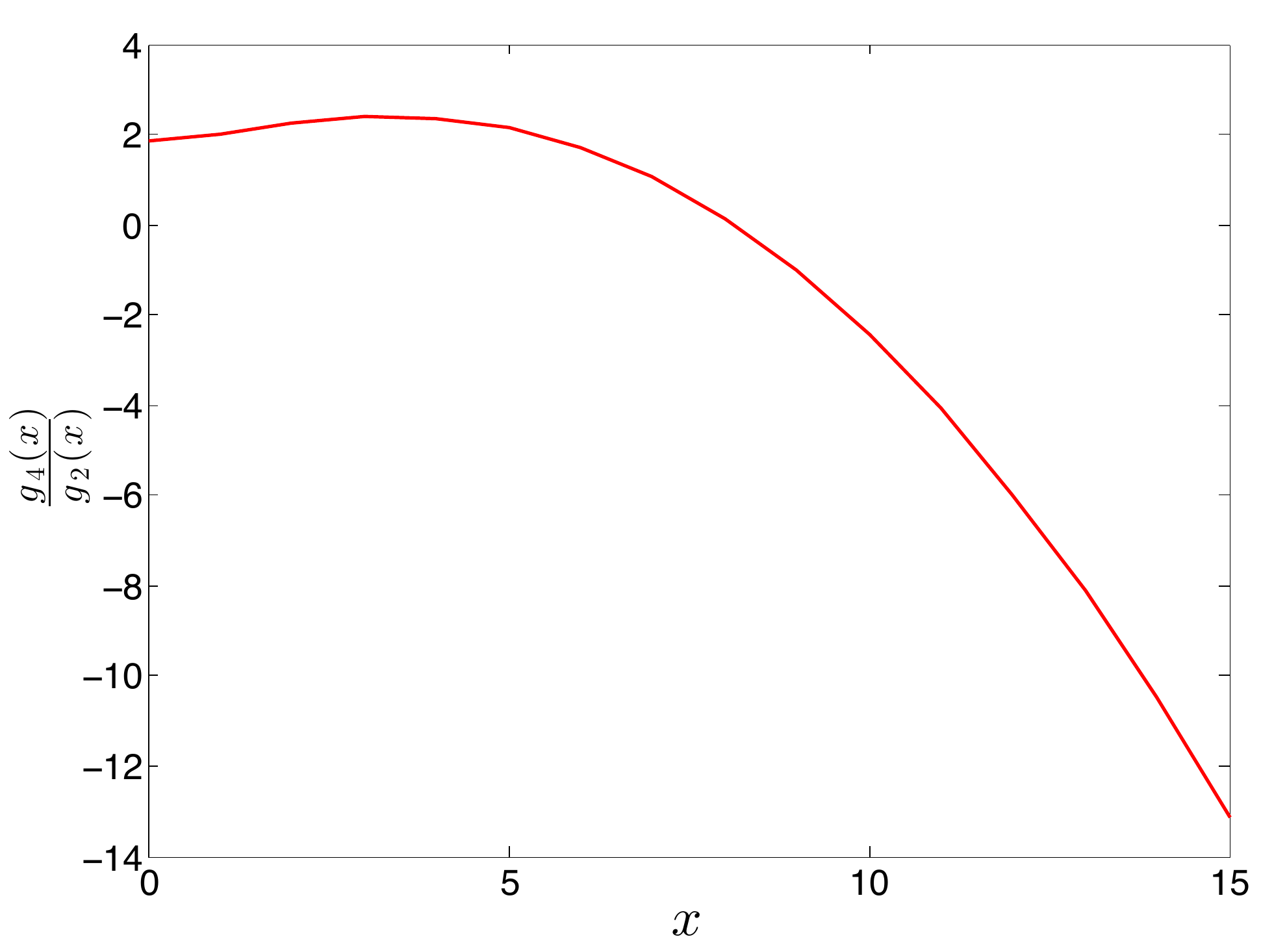}
\caption{\small{$g_4/g_2$ as a function of $x$. Fourth order
correction becomes more relevant when approaching the short wavelength limit.}} \label{fig3}
\end{figure}

\subsection{Nontrivial third order effects}
In the previous example, the third order correction to the Casimir energy
vanished. We will now see that, when the corrugated surface can be
described by a combination of sinusoidal functions of different
wavelengths, a nonvanishing correction may arise to this order,  when the wavelengths 
satisfy a certain condition.  Moreover, unlike the second order correction,
the $O(\eta^3)$ contribution depends on the relative phase of the
sinusoidal components of the corrugation.

Although one could consider a rather general corrugation, for the sake of
definiteness we shall deal with the particular case
\begin{equation}
\eta( {\mathbf x_\parallel})=\epsilon^{(1)} \sin(q_1^{(1)} x_1 +\phi^{(1)})+\epsilon^{(2)} \sin(q_1^{(2)} x_1 +\phi^{(2)})\, .
\end{equation}
The second order correction to the vacuum energy is easily evaluated and gives
\begin{eqnarray}
{\mathcal E^{(2)}_{\rm vac}}&=& \frac{1}{2}\sum_{j=1}^{2}\left(\epsilon^{(j)}\right)^2f^{(2)}(q^{(1)})\nonumber \\ &=&     
-\frac{\pi^2 }{1440 \, a^5}\sum_{j=1}^{2}\left(\epsilon^{(j)}\right)^2 g_2(q_1^{(j)} a),
\end{eqnarray}
where $q^{(1)}=(0,q^{(1)}_1,0,..0)$. This is just the sum of the energies of each Fourier component of the correlation.

In order to evaluate the third order contribution to the Casimir energy, recalling Eq.(\ref{eq:resO3}) we write
\begin{equation}\label{O3}
	{\mathcal E_{\rm vac}^{(3)}} \,=\, {\mathcal E_{\rm vac}^{(3,l)}} \,+\, {\mathcal E_{\rm vac}^{(3,b)}}\,+\, 
{\mathcal E_{\rm vac}^{(3,t)}}\;. 
\end{equation}
The different terms in the above equation can be evaluated  computing the Fourier transform of $\eta( {\mathbf x_\parallel})$ and using Eqs.(\ref{Gamma31}), (\ref{Gamma32}), and (\ref{Gamma33}). Assuming that $q_1^{(2)}>q_1^{(1)}>0$, it is easy to see that the third order doesn't vanish only when $q_1^{(2)}=2 q_1^{(1)}$. In this case we obtain
\begin{eqnarray}
	{\mathcal E_{\rm vac}^{(3,l)}}&=&\frac{\pi^6}{8064\,
a^6}\left(\epsilon^{(1)}\right)^2\epsilon^{(2)}\sin(2\phi^{(1)}-\phi^{(2)})\, ,\\
	{\mathcal E_{\rm vac}^{(3,b)}}&=& \frac{1}{16}\left(\epsilon^{(1)}\right)^2\epsilon^{(2)} 
	\big[ f_b(2q^{(1)}) +  2 f_b(q^{(1)}) \big] \nonumber\\
	&\times& \sin(2\phi^{(1)} - \phi^{(2)})\, ,
\end{eqnarray}
and
\begin{eqnarray}
{\mathcal E_{\rm vac}^{(3,t)}}&= &
	\frac{1}{24}\left(\epsilon^{(1)}\right)^2\epsilon^{(2)} \sin(2\phi^{(1)}-\phi^{(2)})\\
&\times& 
\left[f_t(q^{(1)},q^{(1)})+
f_t(2q^{(1)},-q^{(1)})+
f_t(-q^{(1)},2q^{(1)})\right]\nonumber .
\end{eqnarray}
Unlike the second order corrections, the Casimir energy depends on the phases $\phi^{(j)}$ of the Fourier components of the corrugation.

\section{Conclusions}\label{sec:conc}

In this paper we have extended the perturbative results for the Casimir
energy between slightly deformed mirrors up to the fourth order in the
amplitude of the deformations.  For simplicity we considered the case of a
deformed mirror in front of a plane mirror, for a scalar field satisfying
Dirichlet boundary conditions.  The third and fourth order results are important to improve the accuracy of the 
(almost) analytic  perturbative calculations, to have an explicit way to evaluate the validity of the 
rather simpler second order results, and to provide a benchmark for 
complex numerical calculations.   The results could be generalized, for
example, to the case of the electromagnetic field in a geometry with two deformed mirrors.

As a side point of the perturbative calculations, we have also discussed the relation between different 
approximations usually
considered to compute Casimir forces. For the geometry considered in this
paper, the Casimir energy is a functional of the shape of the surface
$\psi=a+\eta$. This functional can be expanded in powers of $\eta/a$, as we
did here, assuming small deviations from a flat surface.  When the 
additional assumption that the function $\eta$ is slowly varying is
reliable, a resummation  of the  perturbative series is possible, and
approximating the form factors by their zero-momentum values,
the final result coincides with the PFA.  
This is a non perturbative result, valid for arbitrary amplitudes as long
as the surface is gently curved.  The PFA can be improved by expanding
the form factors around $k_\parallel =0$. If this expansion only involves
even powers of the momentum, the higher order corrections can be written in
terms of derivatives of the shape of the surface. This is the derivative
expansion for the Casimir energy we proposed in previous papers
\cite{Fosco:2011xx}. On the other hand, when the form factors contain
nonanalytic terms, the corrections to the PFA will be nonlocal, as it
indeed happens for scalar field satisfying Neumann boundary conditions for 
nonzero temperature~\cite{highT}.  We have also obtained explicit 
expressions for the Casimir energy under the opposite assumption, i.e.
strongly varying surfaces. In this case, the final result could be written in terms of 
nonlocal operators, as $(-\nabla^2)^{1/2}$.

Finally, we have presented some explicit examples. On the one hand, 
we have evaluated numerically the results, up to the fourth
order, for the Casimir energy for the case of a sinusoidal corrugation. 
The numerical results reproduce the analytic results expected in the
limits of low and high wavelengths. The fourth order term is particularly
relevant when evaluating the Casimir energy in the limit of short
wavelengths, where it becomes dominant.    The results presented in the paper
allowed us to  assess the validity of the second order calculations obtained 
in previous works, without the necessity of a full numerical evaluation of the Casimir energy.

On the other hand, we have shown the 
case in which the corrugation consists of a combination of two sinusoidal
functions of different wavelengths, one being twice the other. 
The distinctive characteristic of the result for this case is that, when
relation between the two wavelengths holds, the Casimir
energy bears a dependence on the relative phase of the sinusoidal
functions appearing in the combination.
This `interference« effect was absent in the second order term, namely, the
relative phases of the components are irrelevant.

\section*{Appendix}

In this Appendix we prove that the first terms in the perturbative expansion reproduces the PFA approximation when the form factors are evaluated at zero momentum.
The first term in the series of Eq.(\ref{low}) is given in Eq.(\ref{gamma1}). Indeed, comparing both equations we obtain
\begin{equation}
h^{(1)}(0)=a^4\int \frac{d^3p_\parallel}{(2\pi)^3}
	\, B(p_\parallel) \, |p_\parallel|=\frac{\pi^2}{480}\,  ,
	\end{equation}
that reproduces the linear term in $\eta$ in Eq.(\ref{cons}).

In order to evaluate the term quadratic in $\eta$, we compare  Eqs.(\ref{low}) and (\ref{eq:second}). We get
\begin{equation}
h^{(2)}(0,0)=-a^5\int \frac{d^3p_\parallel}{(2\pi)^3}
	\, B(p_\parallel)(1 +B(p_\parallel)) \, |p_\parallel|^2=-\frac{\pi^2}{240}\,  ,
	\end{equation}
giving the quadratic term in Eq.(\ref{low}).

The zero momentum contribution to the third order can be read from Eqs.(\ref{Gamma31}), (\ref{Gamma32}), and (\ref{Gamma33})
\begin{eqnarray}
h^{(3)}(0,0,0)&=&a^6\int \frac{d^3p_\parallel}{(2\pi)^3}   |p_\parallel|^3\nonumber \\ &\times &
 \left[\frac{B}{6}-\frac{B}{2}+\frac{B}{3}(4B^3+6B+3)\right]\, ,
\end{eqnarray}
where $B=B(p_\parallel)$. The three terms in the integrals are the contributions coming from $\Gamma^{(3,j)}, j=1,2,3$, respectively.
The evaluation of the integral gives $h^{(3)}(0,0,0)=\pi^2/144$, reproducing the third term in Eq.(\ref{cons}).

The evaluation of the fourth order is more cumbersome but straightforward. The form factor  $h^{(4)}(0,0,0,0)$ can be written 
as a linear combination
of $f^{(4,i)},\, i=1,...5$, all of them evaluated at zero momenta. Adding all contributions we obtain
\begin{eqnarray}
h^{(4)}(0,0,0,0)&=&-a^7\int \frac{d^3p_\parallel}{(2\pi)^3}   |p_\parallel|^4 \nonumber \\ &\times & \left[\frac{1}{3}B +\frac{7}{3}B^2+4B^3+2B^4\right]\, .
\end{eqnarray}
Evaluating the integral we obtain $h^{(4)}(0,0,0,0)=-\pi^2/96$, in agreement with the fourth order  term in Eq.(\ref{cons}).

\section*{Acknowledgements}
This work was supported by ANPCyT, CONICET, UBA and UNCuyo.

\end{document}